\documentstyle[iopconf1]{article}
\begin{document}
\title{Baryogenesis through lepton number violation }

\author{Utpal Sarkar\dag\footnote{E-mail:utpal@prl.ernet.in}}

\affil{\dag Physical Research Laboratory, Ahmedabad - 380 009, INDIA}

\beginabstract

The most promising scenarios of baryogenesis seems to be the
one through lepton number violation. Lepton number violation
through a Majarana mass of the right-handed neutrinos
can generate a lepton asymmetry of the universe when the
right-handed neutrinos decay. The
left-handed neutrinos get small Majorana masses through see-saw
mechanism in these models.
A triplet higgs scalar violating lepton number explicitly
through its couplings to two leptons or two higgs doublets
can also naturally give small Majorana masses to the left-handed
neutrinos and also generate a lepton asymmetry of the universe.
We review both these models of leptogenesis, where the lepton number 
asymmetry then gets converted to a baryon asymmetry of the universe
before the electroweak phase transition.

\endabstract 

\section{Introduction} 

To get the baryon asymmetry of the universe \cite{kolb} starting from a
symmetric universe, one requires \cite{sakh} three conditions (A) {\it 
Baryon number violation}, (B) {\it $C$ and $CP$ violation}, and (C) {\it 
Departure from thermal equilibrium}. In grand unified theories (GUTs)
all these conditions are satisfied \cite{gutbar,gutrev}, but the
generated asymmetry conserves $(B-L)$. It was then 
realised that  the chiral nature  of the weak interaction  also
breaks the global baryon  and lepton numbers in the standard model 
\cite{hooft}. At finite
temperature these baryon and lepton number violating interactions were 
found  to be very  strong in the presence of
some static topological field configuration - sphalerons \cite{krs}.
Although the anomalous sphaleron processes conserves $(B-L)$, the
GUT $(B+L)$ asymmetry will be completely washed out by these 
interactions.

Attempts were then made to make use of the baryon number 
violation of the standard model to generate a baryon asymmetry of the 
universe. However, in these models one needs to protect
the generated baryon asymmetry after the phase transition, which 
requires the mass of the standard model doublet higgs boson to be
lighter than the present experimental limit of 95 GeV. 
Then the most interesting
scenario remains for the understanding of the baryon number of the
universe is through lepton number violation \cite{fy1}--\cite{last}, 
which is also referred to as leptogenesis. 

In models of leptogenesis 
one generates a lepton asymmetry of the universe, which is
the same as the $(B-L)$ asymmetry of the universe at some high energy. 
This $(B-L)$ asymmetry of the universe then get converted to the baryon 
asymmetry of the universe during the period when the sphaleron fields 
maintain the baryon number violating interactions in equilibrium. Since
lepton number violation is the source of leptogenesis, they are related
to models of neutrino masses. In this article we shall review two 
scenarios of leptogenesis. In the first scenario right handed neutrinos
are introduced, which gets a Majorana mass and breaks lepton number
\cite{fy1}. The left-handed neutrinos get small Majorana masses
through see-saw mechanism \cite{seesaw}. In the second scenario
only a triplet higgs is introduced and the fermion content of the
standard model is unaltered \cite{ma,ma1,triplet}. Unlike earlier 
treatments, lepton number is now broken explicitly at a very high
scale \cite{ma,ma1}. Although the triplet is very heavy, its vev 
becomes of the order of eV to give very small Majorana mass to the
neutrinos naturally \cite{ma}. Decays of the triplet higgs generate
a lepton asymmetry of the universe at very high scale. 

In the next section we shall discuss the electroweak anomalous 
processes and then how the baryon and lepton numbers of the
universe gets related to the $(B-L)$ number of the universe.
This will imply that if there is vary fast lepton number violation in the
universe during the period when these processes are in equilibrium, 
that can also wash out the baryon asymmetry
of the universe \cite{fy2,lgb}. In the following two sections we present
the two scenarios of leptogenesis.

\section{Sphaleron processes in thermal equilibrium and 
relation between baryon and lepton numbers}

Anomaly breaks any classical symmetry of the lagrangian at the 
quantum level. So, all local gauge theories should be free of
anomalies. However, there may be anomalies corresponding to any
global current. That will simply mean that such global symmetries
of the classical lagrangian are broken through quantum effects.

In the standard model the chiral nature of the weak interaction
makes the baryon and lepton number anomalous and give
us non vanishing axial current \cite{hooft}
$$ \delta_\mu j^{\mu 5}_{(B+L)} = 6 [ {\alpha_2 \over 8 \pi}
W_a^{\mu \nu} \tilde{W}_{a \mu \nu} +  {\alpha_1 \over 8 \pi}
Y^{\mu \nu} \tilde{Y}_{\mu \nu} ] $$
which will break the $(B+L)$ symmetry, while still preserving $(B-L)$,
during the electroweak phase transition, 
$$ \Delta (B+L) = 2 N_g {\alpha_2 \over 8 \pi} \int d^4 x
W_a^{\mu \nu} \tilde{W}_{a \mu \nu} = 2 N_g \nu  $$
But their rate is very
small at zero temperature, since they are suppressed by quantum 
tunnelling probability, $\exp[- {2 \pi \over \alpha_2} \nu] ,$ where
$\nu$ is the Chern-Simmons number.

At finite temperature, however, it has been shown that there exists
non-trivial static topological soliton configuration, 
called the sphalerons, which enhances the baryon number violating 
transition rate \cite{krs} and the suppression
factor is now replaced by the Boltzmann factor $ \exp[- {V_0 \over
T } \nu]$ where the potential or the free energy $V_0$
is related to the mass of the sphaleron field, which is 
about TeV. As a result, at temperatures between 
\begin{equation} 
10^{12} GeV > T > 10^2 GeV  \label{per}
\end{equation}
the sphaleron mediated baryon and lepton number violating
processes are in equilibrium. For the simplest scenario of $\nu = 1$,
the sphaleron induced processes are $\Delta B = \Delta L = 3$, 
given by,
\begin{equation} 
|vac> \longrightarrow [u_L u_L d_L e^-_L + c_L c_L s_L \mu^-_L
+ t_L t_L b_L \tau^-_L] . \label{sph}
\end{equation}
These baryon and lepton number violating fast processes will wash
out any pre-existing baryon or lepton number asymmetry, or will
convert any pre-existing $(B-L)$ asymmetry of the 
universe to a baryon asymmetry of the universe, 
which can be seen from an analysis of the
chemical potential \cite{ht}. 

We consider all the particles to be 
ultrarelativistic and ignore small mass corrections. The
particle asymmetry, {\it  i.e.} the difference  between the number of
particles ($n_{+}$) and the number of antiparticles ($n_{-}$) can be
given in terms of the chemical potential of the particle species $\mu$ 
(for antiparticles the chemical potential is $-\mu $) as
$n_{+}-n_{-}=n_{d}{\frac{gT^{3}}{6}}\left( {\frac{\mu}{T}}\right), $
where $n_{d}=2$ for bosons  and $n_{d}=1$ for fermions. 

In the standard model there are quarks and leptons $q_{iL}, u_{iR},
d_{iR}, l_{iL}$ and $e_{iR}$; 
where, $i = 1,2,3$ corresponds to three generations. 
In addition, the scalar sector consists of the usual Higgs doublet
$\phi$,
which breaks the electroweak gauge symmetry $SU(2)_L \times U(1)_Y$
down to $U(1)_{em}$. 
In  Table 1, we presented the relevant 
interactions and the corresponding relations between  
the chemical potentials.  In the third column we give the chemical potential 
which we eliminate using the given relation.  We start with chemical 
potentials of all the quarks ($\mu _{uL},\mu _{dL},\mu _{uR},\mu _{dR}$); 
leptons ($\mu _{aL},\mu _{\nu aL},\mu _{aR}$,  where 
$ a=e,\mu,\tau $); gauge bosons ($\mu _{W}$  for $W^{-}$, 
and 0 for all others); and the Higgs scalars ($\mu _{-}^{\phi  },
\mu _{0}^{\phi}$).  

\begin{table}[htb]
\caption {Relations among the chemical potentials}
\begin{center}
\begin{tabular}{||c|c|c||}
\hline \hline
Interactions& $\mu$ relations&$\mu $ eliminated \\
\hline
{$D_{\mu }\phi ^{\dagger}D_{\mu
}\phi $}&{$\mu _{W}=\mu _{-}^{\phi }+\mu _{0}^{\phi
}$}&{$\mu_{-}^{\phi }$}\\
{$\overline{q_{L}}\gamma _{\mu}q_{L}W^{\mu }$}&{$\mu
_{dL}=\mu _{uL}+\mu _{W}$}&{$\mu_{dL}^{{}}$}\\
{$\overline{l_{L}}\gamma _{\mu }l_{L}W^{\mu
}$}&{$\mu_{iL}^{{}}=\mu _{\nu iL}^{{}}+\mu
_{W}$}&{$\mu_{iL}$}\\
{$\overline{q_{L}}u_{R}\phi ^{\dagger
}$}&{$\mu_{uR}=\mu _{0}+\mu_{uL}$}&{$\mu_{uR}^{{}}$}\\
{$\overline{q_{L}}d_{R}\phi $}&{$\mu
_{dR}=-\mu_{0}+\mu _{dL}$}&{$\mu_{dR}$}\\
{$\overline{l_{iL}}e_{iR}\phi
$}&{$\mu _{iR}^{{}}=-\mu_{0}+\mu _{iL}^{{}}$}&{$\mu_{iR}$}\\
\hline \hline
\end{tabular}
\end{center}
\end{table}

The chemical potentials of the neutrinos always enter as a sum 
and for that reason we can consider it as one parameter.
We can then express all the chemical potentials in terms of the following
independent chemical potentials only,
$ \mu _{0}=\mu _{0}^{\phi };~~\mu _{W};~~\mu _{u}=\mu _{uL};~~
\mu = \sum_i \mu _{i}= \sum_i \mu _{\nu iL} $. 
We can further eliminate one of these four potentials by making use of the
relation given by the sphaleron processes, $
3\mu _{u}+2\mu _{W}+ \mu =0 $.
We then express the baryon number, lepton numbers and the
electric charge and the hypercharge number densities in terms of these
independent chemical potentials,
\begin{eqnarray}
&&B =12\mu _{u}+6\mu _{W} ; \hskip 1.6in
L_{i} =3\mu +2\mu _{W}-\mu _{0}  \hskip .4in \nonumber \\
&&Q =24 \mu _{u}+(12+2m)\mu _{0}-(4+2m)\mu _{W} ; \hskip .2in
Q_{3} =-(10+m)\mu _{W} \hskip .4in \nonumber
\end{eqnarray}
where $m$ is the number of Higgs doublets $\phi$.

At temperatures above the electroweak phase transition, $T>T_{c}$, both 
$<Q>$ and $<Q_{3}>$ must vanish, while below the critical temperature
$<Q>$ should vanish, but since $SU(2)_L$
is now broken we can consider $\mu_0^\phi =0$ and $Q_3 \neq 0$. 
These conditions and the
sphaleron induced $B-L$ conserving, $B+L$ violating condition will
allow us to write down the baryon asymmetry 
in terms of the $B-L$ number density as,
\begin{equation}
B(T>T_c) = \frac{24+4m}{66+13m}~(B-L) \hskip .3in
B(T<T_c) = \frac{32+4m}{98+13m}~(B-L).
\end{equation}
Thus the baryon and lepton number asymmetry of the universe after 
the electroweak phase transition will depend on the primordial $(B-L)$ 
asymmetry of the universe, while all the primordial $(B+L)$ asymmetry
will be washed out. 

\section{Leptogenesis with right-handed neutrinos}

To give a small Majorana mass to the left-handed neutrino, right-handed
neutrinos were introduced. Although it is most natural to introduce
a right handed neutrino in left-right symmetric models \cite{lr1,lr2},
in the minimal scenario the standard model is extended with right
handed neutrinos ($N_{Ri},i=e,\mu,\tau$). In these models  
neutrino masses come from the see-saw mechanism \cite{seesaw}. 
The lagrangian for the lepton sector containing the 
mass terms of the singlet right handed neutrinos $N_i$ and the
Yukawa couplings of these fields with the light leptons is
given by,
\begin{equation}
{\cal L}_{int} = h_{\alpha i}~  \overline{\ell_{L \alpha}} \phi ~N_{Ri}
 + M_i ~\overline{(N_{Ri})^c}~ N_{Ri}
\end{equation}
where, $\ell_{L \alpha}$  are the
light leptons,  $h_{\alpha i}$ are the complex  Yukawa  couplings
and $\alpha$ is the generation  index.  
Without  loss  of  generality  we work in a basis  in  which  the
Majorana mass matrix of the right handed neutrinos is real and 
diagonal with eigenvalues $M_i$, and assume $M_3 > M_2 > M_1$.

Because of the Majorana mass term, 
the decay of $N_{Ri}$ into a lepton and an antilepton,
\begin{eqnarray}
  N_{Ri}  &\to& \ell_{jL} + \bar{\phi}, \nonumber \\
   &\to&  {\ell_{jL}}^c + {\phi} .\label{N}
\end{eqnarray}
breaks lepton number, which can generate a lepton asymmetry
of the universe. 
There are two sources of CP violation in this scenario :

\begin{figure}[hb]
\vskip 2in\relax\noindent\hskip -.3in\relax{\includegraphics{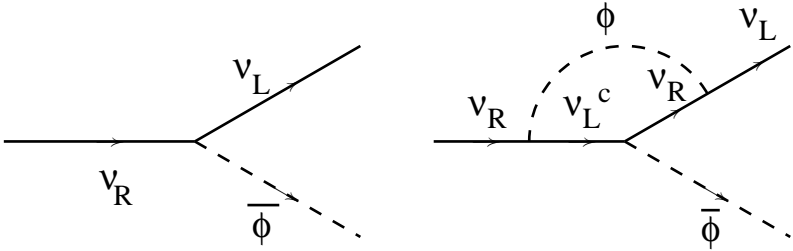}}
\caption{Tree and one loop vertex correction 
diagrams contributing to the generation of lepton asymmetry
in models with right handed neutrinos}
\end{figure}

\begin{figure}[t]
\vskip 2.25in\relax\noindent\hskip -.3in\relax{\includegraphics{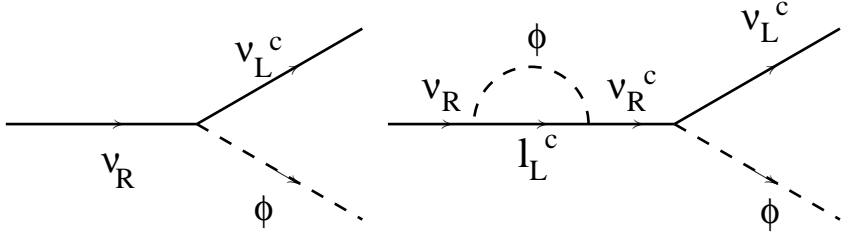}}
\caption{Tree and one loop self energy 
diagrams contributing to the generation of lepton asymmetry in models
with right handed neutrinos}
\end{figure}

\begin{itemize}
\item[$(i)$] vertex type diagrams which interferes with the tree level
diagram given by figure 2. This is similar to the $CP$ violation
coming from the penguin diagram in $K-$decays.

\item[$(ii)$] self energy diagrams could interfere with the tree level
diagrams to produce CP violation as shown in figure 3. This is 
similar to the $CP$ violation in $K-\bar{K}$ oscillation, entering
in the mass matrix of the heavy Majorana neutrinos.

\end{itemize}

In the first paper on leptogenesis \cite{fy1}, the vertex type
diagram was only mentioned. Subsequently, it has been extensively 
studied \cite{vertex} and the amount of $CP$ asymmetry is calculated
to be,
\begin{equation}
\delta  = - {1 \over 8 \pi} \frac{M_1 M_2}{M_2^2 - M_1^2}
\frac{{\rm Im} [ \sum_\alpha (h_{\alpha 1}^\ast h_{\alpha 2})
 \sum_\beta (h_{\beta 1}^\ast h_{\beta 2}) ] }{ \sum_\alpha 
|h_{\alpha 1}|^2}
\end{equation}
In this expression it has been assumed that the main contribution
to the asymmetry comes from the lightest right handed neutrino 
($N_1$) decay, when the other heavy neutrinos have already decayed away.

The heavy neutrinos decay into light leptons and higgs doublets. 
Because of $C$ and $CP$ violation, the decays of $N_{1R}$ would 
produce more anti-leptons than leptons. This will be compensated
by an equal amount of asymmetry in phi, so that there is no charge
asymmetry. 

Initially the self energy diagram was considered for $CP$ violation 
as an additional contribution \cite{self}. It was then pointed out 
\cite{pas1} that this $CP$ violation enters in the mass matrix as
in the $K-\bar{K}$ oscillation. Before they decay, the right handed
neutrinos were considered to oscillate to an anti-neutrino and since
the rate of $particle \to anti-particle \neq anti-particle \to particle$,
an asymmetry in the right handed neutrino was obtained before they
decay. As a result, when the two heavy 
right handed neutrinos are almost degenerate, {\it i.e.}, the mass
difference is comparable to their width, there may be a resonance 
effect which can enhance the $CP$ asymmetry by few orders of magnitude
\cite{pas2}. This effect was then confirmed by other calculations
\cite{deg1,deg2}. Ref \cite{deg1} gives a very rigorous treatment
based on a field-theoretic resummation 
approach used earlier to treat unstable intermediate
states, which was used earlier in different contexts \cite{pil}.
This issue has been reviewed in another talk in this meeting
\cite{pil1}. 
 
When the mass difference is large compared to the width, the $CP$
asymmetry generated though the mixing of the heavy neutrinos is same
as the vertex correction. 
These two contributions add up to produce the final lepton asymmetry 
of the universe.

Although the $CP$ asymmetry was found to be non-vanishing,
in thermal equilibrium unitarity and $CPT$ would mean that there is
no asymmetry in the final decay product. However, when the 
out-of-equilibrium condition of the heavy neutrinos decay is 
considered properly, one could get an asymmetry as expected. 
Consider the decays of $K_L$ and
$K_S$. If they were generated in the early universe, 
in a short time scale $K_S$ could decay and 
recombine, but $K_L$ may not be able to decay or recombine. As a result in
the decay product there will be an asymmetry in $K$ and $\bar{K}$
if there is $CP$ violation. In the lepton number violating two 
body scattering processes $CP$ violation
in the real intermediate state plays the most crucial role \cite{last},
which comes since the decay take place away from thermal equilibrium.

Whether a system is in equilibrium or not can be understood by solving
the Boltzmann equations \cite{fry}. 
But a crude way to put the out-of-equilibrium
condition is to say that the universe expands faster than some 
interaction rate. This may be stated as
\begin{equation}
\Gamma < 1.7 \sqrt{g_*} {T^2 \over M_P}
\end{equation}
where, $\Gamma$ is the interaction rate under discussion, $g_*$ is the 
effective number of degrees of freedom available at that temperature $T$, 
and $M_P$ is the Planck scale.

In the case of right handed neutrino decay, the asymmetry is generated
when the lightest one (say $N_1$) decay. Before its decay, the pre-existing
lepton asymmetry is washed out by its lepton number violating interactions.
So the out-of-equilibrium condition now implies that the lightest 
right-handed neutrino should satisfy the out-of-equilibrium condition 
when it decays, which is given by,
\begin{equation}
{|h_{\alpha 1}|^2 \over 16 \pi} M_1 < 1.7 \sqrt{g_*} {T^2 \over M_P}
\hskip .5in {\rm at}~~T = M_1
\end{equation}
which gives a bound on the mass of the lightest right-handed 
neutrino to be $ m_{N_1} <  10^{7} GeV . $ Finally the lepton asymmetry
and hence a $(B-L)$ asymmetry 
generated at this scale gets converted to a baryon asymmetry of the
universe in the presence of sphaleron induced processes.

\section{Leptogenesis with triplet higgs}

There are several alternative scenarios to give a small mass to the 
left-handed neutrinos \cite{ma,ma1,triplet,oth1,oth2}. However, at present
lepton asymmetry could be generated only in models with triplet higgs
\cite{ma}. In this scenario \cite{ma} one adds two complex 
$SU(2)_L$ triplet higgs scalars ($\xi_a \equiv (1,3,-1); a = 1,2$). 
The $vev$s of the triplet higgses can give small
Majorana masses to the neutrinos \cite{ma,ma1,triplet}
through the interaction
\begin{equation}
f_{ij} [\xi^0 \nu_i \nu_j + \xi^+ (\nu_i l_j + l_i \nu_j)/\sqrt 2 
+ \xi^{++} l_i l_j] + h.c.
\end{equation}
If the triplet higgs acquires a $vev$ and break lepton number spontaneously, 
then there will be Majorons in the problem which is ruled out by precision
Z--width measurement at LEP. However, in a variant of this
model \cite{ma} lepton number is broken explicitly through an interaction
of the triplet with the higgs doublet
\begin{eqnarray}
V &=&  \mu (\bar \xi^0 \phi^0 \phi^0 + \sqrt 2 \xi^- \phi^+ \phi^0 + \xi^{--} 
\phi^+ \phi^+) + h.c.
\end{eqnarray}
Let $\langle \phi^0 \rangle = v$ and $\langle \xi^0 \rangle = u$, then the 
conditions for the minimum of the potential relates the $vev$ of the 
two scalars by
$ u \simeq {{-\mu v^2} \over M^2}, \label{min} $,
where $M$ is the mass of the triplet higgs scalar and 
the neutrino mass matrix becomes $-2 f_{ij} \mu v^2 / M^2 = 2 f_{ij} u$.

In this case the lepton number violation comes from the
decays of the triplet higgs $\xi_a$, 
\begin{equation}
\xi_a^{++} \rightarrow \left\{ \begin{array} {l@{\quad}l} l_i^+ l_j^+ & 
(L = -2) \\ \phi^+ \phi^+ & (L = 0) \end{array} \right.
\end{equation}
The coexistence of the above two types of final states indicates the 
nonconservation of lepton number.  On the other hand, any lepton asymmetry 
generated by $\xi_a^{++}$ would be neutralized by the decays of $\xi_a^{--}$, 
unless CP conservation is also violated and the decays are out of thermal 
equilibrium in the early universe. In this case there are no
vertex corrections which can introduce CP violation. The only source
of CP violation is the self energy diagrams of figure 4. 

\begin{figure}[t]
\vskip 2.5in\relax\noindent\hskip -.5in\relax{\includegraphics{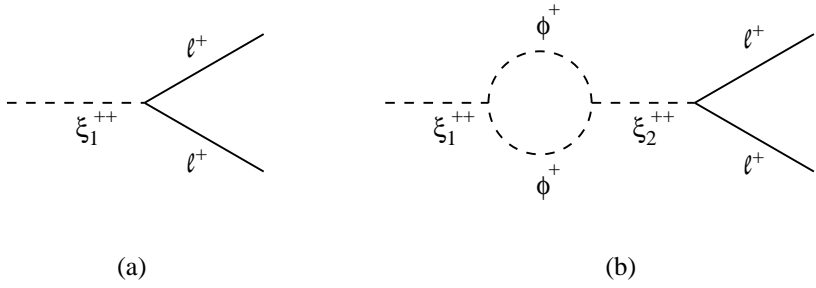}}
\caption{The decay of $\xi_1^{++} \to l^+ l^+$ at tree level (a) and in 
one-loop order (b).  A lepton asymmetry is generated by their interference
in the triplet higgs model for neutrino masses.}
\end{figure}

If there is only one $\xi$, then the 
relative phase between any $f_{ij}$ and $\mu$ can be chosen real.  Hence 
a lepton asymmetry cannot be generated.  With two $\xi$'s, even if there is 
only one lepton family, one relative phase must remain.  
As for the possible relative phases 
among the $f_{ij}$'s, they cannot generate a lepton asymmetry because they 
all refer to final states of the same lepton number.

In the presence of the one loop diagram, the mass matrix ${M_a}^2$ and
${M_a^*}^2$ becomes different. This implies that the rate of 
$\xi_b \to \xi_a$ no longer remains to be same as $\xi_b^* \to \xi_a^*$.
Since by $CPT$ theorem $\xi_b^* \to \xi_a^* \equiv \xi_a \to \xi_b$, 
what it means is that now $ \Gamma[\xi_a \to \xi_b] \neq
\Gamma[\xi_b \to \xi_a] .$
This is a different kind of CP violation compared to the CP violation
in models with right handed neutrinos. If we consider that the $\xi_2$
is heavier than $\xi_1$, then after $\xi_2$ decayed out the decay of
$\xi_1$ will generate an lepton asymmetry given by,
\begin{equation}
\delta \simeq 
{{Im \left[ \mu_1 \mu_2^* \sum_{k,l} f_{1kl} f_{2kl}^* \right]} \over 
{8 \pi^2 (M_1^2 - M_2^2)}} \left[ {{ M_1} \over 
\Gamma_1}  \right].
\end{equation}
In this model the out-of-equilibrium condition is satisfied when
the masses of the triplet higgs scalars are heavier than $10^{13}$ GeV.

The lepton asymmetry thus generated after the Higgs triplets 
decayed away would be the same as the $(B-L)$ asymemtry before the
electroweak phase transition. During the electroweak phase 
transition, the presence of sphaleron fields would relate
this $(B-L)$ asymmetry to the baryon asymmetry of the universe. The
final baryon asymmetry thus generated can then be given by the
approximate relation $
{n_B \over s} \sim {\delta_2 \over 3 g_* K ({\rm ln} K)^{0.6}} $
To obtain a neutrino mass of order eV or less, as well as the observed baryon 
asymmetry of the universe, we may choose $M_2 = 10^{13}$ GeV, $\mu_2 = 
2 \times 10^{12}$ GeV, and $f_{233} \sim 1$, then $m_{\nu_\tau} \sim 1$ eV, 
assuming that the $M_1$ contribution is negligible.  Now let $M_1 = 3 \times 
10^{13}$ GeV, $\mu_1 = 10^{13}$ GeV, and $f_{1kl} \sim 0.1$, then the decay of 
$\psi_2^\pm$ generates a lepton asymmetry $\delta_2$ of about $8 \times 
10^{-4}$ if the CP phase is maximum.  Using $M_{Pl} \sim 10^{19}$ GeV and $g_* 
\sim 10^2$, we find $K \sim 2.4 \times 10^{3}$.  Hence $n_B/s \sim 10^{-10}$ 
as desired.

\section{Summary }

There are several models of neutrino masses which require lepton 
number violation. In models with right handed neutrinos, where the
left-handed neutrinos get a see-saw mass, lepton number violation
is introduced by the Majorana mass term of the right handed neutrinos.
In these models the decays of the right handed neutrinos can generate
a letpon asymmetry of the universe, which can then get converted to
a baryon asymmetry of the universe during the period when the 
sphaleron induced $(B+L)$ violating processes are in equilibrium. 
Lepton asymmetry of the universe may also be generated in models
with triplet higgs scalars. In these models lepton number is violated
explicitly through the coupling of the triplet higgs at very high
energy. However, these triplet higgs scalars get a very tiny $vev$
through see-saw mechanism in the higgs sector and can naturally
produce light left-handed Majorana neutrinos without introducing 
any right-handed neutrinos. In this model the decay of the triplet
higgs can generate a lepton asymmetry of the universe at a very 
high energy, which can then get converted to a baryon asymmetry of
the universe. At present we cannot distinguish these two equivalent models
of neutrino masses and leptogenesis with a right handed neutrino 
or with a triplet higgs scalar from each other.  

\section*{Acknowledgement}

I would like to thank the organisers of the Dark Matter meeting at
Heidelberg for fantastic arrangements and hospitality and
acknowledge a financial support from the Alexander von Humboldt 
Foundation to participate in this meeting.


\vspace{-14pt}
\begin{thebibliography}{99}

\bibitem{kolb}  Kolb E W and  Turner M S 1989,  {\it  The  Early
  Universe} (Addison-Wesley, Reading, MA).

\bibitem{sakh} Sakharov A D 1967, {\em Pis'ma Zh. Eksp. Teor. Fiz.} {\bf 5} 32.

\bibitem{gutbar} Yoshimura M 1978, \PRL {\bf 41} 281; 
  E 1979: {\it ibid.} {\bf 42} 7461;

\bibitem{gutrev} Mohapatra R N 1992, {\it Unification and Supersymmetry}
  (Springer-Verlag); 
  \newref Zee A 1982, (ed.) {\it Unity of Forces in the Universe} {\bf 1}
  (World Scientific).
  
\bibitem{hooft} 't Hooft G 1976, \PRL {\bf 37} 8.

\bibitem{krs}  Kuzmin V, Rubakov V and Shaposhnikov M 1985,
  \PL {\bf B 155} 36.

\bibitem{fy1} Fukugita M and Yanagida T 1986, \PL {\bf B 174} 45.

\bibitem{vertex} Langacker P, Peccei R D and Yanagida T 1986,
  {\it Mod. Phys. Lett.} {\bf A 1} 541; 
  \newref Luty M A 1992, \PR {\bf  D 45} 445; 
  \newref Mohapatra R N and Zhang X 1992, \PR {\bf D 45} 5331; 
  \newref Enqvist K and Vilja I 1993, \PL {\bf B 299} 281; 
  \newref Murayama H, Suzuki H, Yanagida T and 
  Yokoyama J 1993, \PRL {\bf 70} 1912; 
  \newref Acker A,  Kikuchi H,  Ma E and Sarkar U 1993,  \PR {\bf D 48} 5006; 
  \newref  O'Donnell P J and Sarkar U 1994, \PR  {\bf D 49} 2118;
  \newref Buchm\"{u}ller W and Pl\"{u}macher M 1996, \PL {\bf B 389} 73;
  \newref Covi L, Roulet E and Vissani F 1996, \PL {\bf B 384} 169;
  \newref  Ganguly A, Parikh J C and Sarkar U 1996, \PL {\bf B 385} 175; 
  \newref Pl\"{u}macher M 1997, \ZP {\bf C 74} 549; 
  \newref Faridani J, Lola S, O'Donnell P J and Sarkar U 1998, hep-ph/9804261.

\bibitem{self} Ignatev A, Kuzmin V and Shaposhnikov M 1979, {\it JETP
  Lett.} {\bf 30} 688;
  \newref Botella F J and Roldan J 1991, \PR  {\bf D 44} 966.
  \newref Liu J and Segre G 1993, \PR {\bf D 48} 4609.
 
\bibitem{pas1} Flanz M, Paschos E A, and Sarkar U 1995, \PL {\bf B 345} 248.

\bibitem{pas2} Flanz M, Paschos E A, Sarkar U and Weiss J 1996, 
  \PL {\bf B 389} 693. 

\bibitem{deg1} Pilaftsis A 1997, \PR {\bf D 56} 5431.

\bibitem{deg2}  Covi L and Roulet E 1997, \PL {\bf B 399} 113.

\bibitem{last} Covi L, Roulet E and Vissani F 1998, \PL {\bf B 424} 101; 
  \newref Buchm\"{u}ller W and Pl\"{u}macher M 1997, hep-ph/9710460 (revised);
  \newref Flanz M and Paschos E A 1998, hep-ph/9805427;
  \newref Rangarajan R, Sarkar U and Vaidya R 1998, hep-ph/9809304.

\bibitem{seesaw}  Gell-Mann M,  Ramond P and Slansky R 1979, in {\it
  Supergravity},  Proceedings  of the Workshop,  Stony Brook, New
  York, 1979,  ed. by P.  van  Nieuwenhuizen  and D.  Freedman
  (North-Holland, Amsterdam); \newref  Yanagida T 1979, in {\it Proc of
  the  Workshop  on Unified  Theories  and  Baryon  Number in the
  Universe},  Tsukuba,  Japan,  edited by A.  Sawada and A.
  Sugamoto (KEK Report No.  79-18, Tsukuba);
  \newref  Mohapatra R N and Senjanovi\'{c} G 1980, \PRL {\bf 44} 912. 

\bibitem{ma} Ma E and Sarkar U 1998, \PRL {\bf 80} 5716.

\bibitem{ma1} Lazarides G and Shafi Q 1998, report no hep-ph/9803397; 
  \newref Ma E 1998, \PRL {\bf 81} 1171;
  \newref Ma E and Sarkar U 1998, hep-ph/9807307;
  \newref Sarkar U 1998, hep-ph/9807466.

\bibitem{triplet} Gelmini G B and Roncadelli M 1981, \PL {\bf B 99} 411;
  \newref Wetterich C 1981, \NP {\bf B 187} 343;
  \newref Lazarides G, Shafi Q and Wetterich C 1981, \NP {\bf B 181} 287;
  \newref Mohapatra R N and Senjanovic G 1981, \PR {\bf D 23} 165;
  \newref Holman R, Lazarides G and Shafi Q 1983, \PR {\bf D 27} 995.

\bibitem{fy2}  Fukugita M and Yanagida T 1990, \PR {\bf D 42} 1285;  
  \newref Barr S M and Nelson A E 1991,  \PL {\bf B 246} 141. 

\bibitem{lgb}  Fischler W,  Giudice G,  Leigh R and 
  Paban S 1991, \PL  {\bf B 258} 45; \newref  Buchm\"{u}ller W
  and Yanagida T 1993, \PL {\bf B 302} 240;
  \newref Dreiner H and Ross G G 1993,  \NP {\bf B 410} 188; 
  \newref  Ilakovac A and Pilaftsis A 1995, \NP {\bf B 437} 491.
  \newref Sarkar U 1997, \PL {\bf B 390} 97.
  \newref Campbell B, Davidson S, Ellis J and Olive K 1991,
   \PL {\bf B 256} 457; \newref Sarkar U 1998, hep-ph/9809209.

\bibitem{ht} Khlebnikov S Yu and Shaposhnikov M E 1988, \NP {\bf B 308} 885; 
  \newref Harvey J A and Turner M S 1990, \PR {\bf D 42} 3344.

\bibitem{lr1} Pati J C and Salam A 1974,  \PR  {\bf  D 10} 275; 
  \newref  Mohapatra R N and Pati J C 1975, \PR {\bf D 11} 566;  
  \newref Mohapatra R N and Senjanovic G 1975, \PR {\bf D 12} 1502;
  \newref  Marshak R E  and Mohapatra R N 1980, \PRL {\bf 44} 1316.

\bibitem{lr2} Pati J C, Salam A and Sarkar U 1983, \PL {\bf B 133} 330.

\bibitem{pil} Papavassiliou J and Pilaftsis A 1995, \PRL {\bf 75}
  3060; 1996 \PR {\bf D 53} 2128; 1996 \PR {\bf D 54} 5315;
  \newref Pilaftsis A 1996, \PRL {\bf 77} 4996;
  1997 \NP {\bf B 504} 61; 1990 \ZP {\bf C 47} 95;
  \newref Pilaftsis A and Nowakowski M 1990, \PL {\bf B 245} 185;
  1991 {\em Mod. Phys. Lett.} {\bf A 6} 1933.

\bibitem{pil1} Pilaftsis A  1998, hep-ph/9810211.

\bibitem{fry} Fry J N, Olive K A and Turner M S 1980, \PRL {\bf 45} 2074; 
  1980 \PR {\bf D 22} 2953; 1980 \PR {\bf D 22} 2977;
  \newref Kolb E W and Wolfram S 1980, \NP {\bf B 172} 224.

\bibitem{oth1}  Nandi S and Sarkar U 1986,  \PRL {\bf 56} 564; 
  \newref Joshipura A S and Sarkar U 1986, \PRL
  {\bf  57} 33; \newref  Masiero A, 
  Nanopoulos D V and Sanda A I 1986, \PRL {\bf 57} 
  663; \newref  Mann R B and Sarkar U 1988,  {\it Int.  Jour.  Mod.
  Phys.}  {\bf A 3} 2165; 

\bibitem{oth2}  Farhi E and Susskind L 1979, \PR {\bf D 20} 3404;
  \newref  Dimopoulos S 1980, \NP {\bf B 168} 69;
 \newref   Zee A 1980, \PL  {\bf B 93} 389; 
 \newref  Wolfenstein L 1980, \NP  {\bf B 175} 93;
 \newref Nussinov S 1985, \PL {\bf B 165} 55.

\end{thebibliography}
\end{document}